\documentclass[12pt]{article}
\usepackage{enumerate}
\usepackage{graphicx}
\usepackage{dcolumn}
\usepackage{bm}
\usepackage{color} 
\usepackage{amssymb} 
\usepackage{amsmath}
\usepackage[mathscr]{euscript}
\usepackage{float}
\restylefloat{table}
\usepackage{authblk}

\newcommand{\be}{\begin{equation}}
\newcommand{\ee}{\end{equation}}
\newcommand{\bearr}{\begin{eqnarray}}
\newcommand{\eearr}{\end{eqnarray}}
\newcommand{\tvE}{\tilde{\vec{E}}}
\newcommand{\tvB}{\tilde{\vec{B}}}
\newcommand{\tvJ}{\tilde{\vec{J}}}

\newcommand{\sS}{\mathscr{S}}

\title{On the Numerical Dispersion of Electromagnetic Particle-In-Cell Code : Finite Grid Instability}
\author[1,2]{M. D. Meyers\thanks{mdmeyers@ucla.edu}} 
\author[1]{C.-K. Huang\thanks{huangck@lanl.gov, corresponding author.}}
\author[1]{Y. Zeng}
\author[1]{S. A. Yi}
\author[1]{B. J. Albright}
\affil[1]{Los Alamos National Laboratory, Los Alamos, New Mexico 87545, USA}
\affil[2]{Department of Physics and Astronomy, University of California Los Angeles, Los Angeles, California 90095, USA}

\begin{document}
\maketitle

\begin{abstract}
The Particle-In-Cell (PIC) method is widely used in relativistic particle beam and laser plasma modeling. However, the PIC method exhibits numerical instabilities that can render unphysical simulation results or even destroy the simulation. For electromagnetic relativistic beam and plasma modeling, the most relevant numerical instabilities are the finite grid instability and the numerical Cherenkov instability. We review the numerical dispersion relation of the electromagnetic PIC algorithm to analyze the origin of these instabilities. We rigorously derive the faithful 3D numerical dispersion of the PIC algorithm, and then specialize to the Yee FDTD scheme. In particular, we account for the manner in which the PIC algorithm updates and samples the fields and distribution function. Temporal and spatial phase factors from solving Maxwell's equations on the Yee grid with the leapfrog scheme are also explicitly accounted for. Numerical solutions to the electrostatic-like modes in the 1D dispersion relation for a cold drifting plasma are obtained for parameters of interest. In the succeeding analysis, we investigate how the finite grid instability arises from the interaction of the numerical 1D modes admitted in the system and their aliases. The most significant interaction is due critically to the correct represenation of the operators in the dispersion relation. We obtain a simple analytic expression for the peak growth rate due to this interaction.
 
\end{abstract}

\section {Introduction}

The Particle-In-Cell (PIC) method \cite{Birdsall-Langdon, Hockney-Eastwood, Dawson1983} is widely-used in numerical simulations of the kinetic behaviors of plasmas and particle beams. The PIC algorithm is a combination of the Lagrangian model for the particles, in which particles can freely move in the simulation domain (subject to boundary conditions) and the grid-based model for the electromagnetic fields, where the fields are solved on the grids from the charge and current of the particles evaluated onto the grids. Contrary to the difference between the particle's continuous spatial variable and the fields' discrete spatial variable, which requires gather/scatter operations in each simulation step, all particle and field quantities in the PIC algorithm are defined on a discrete time variable, albeit with necessary constant time offsets between these quantities due to the staggering of the temporal grid (similarly, there is spatial staggering). As a result, no equivalent gather/scatter operation in time is necessary. Typical Electromagnetic PIC code (referred to as the standard E.M. PIC hereafter) adopts the Yee \cite{Yee1966} Finite Difference Time Domain (FDTD) scheme for solving the Maxwell's equations. The Yee FDTD scheme (and other similar FDTD schemes) by itself, when applied to a simulation of simple medium, is stable as long as the Courant condition is satisfied. However, its coupling with the continuous Lagrangian particle model in PIC is susceptible to various numerical instabilities --- some are encumbering but generally benign, such as the grid heating \cite{Langdon1970}, but many others are very virulent. Recent examples of the latter being the numerical instability found in an relativistic drifting cold plasma for the simulation of the laser wakefield accelerator \cite{Vay2011} and collisionless shock  \cite{Spitkovsky2008, SamueldeFreitasMartins}. Numerical instability also inflates the inherent noise level in a realistic PIC simulation, potentially leading to ``spurious physics", i.e., unphysical numerical artifacts. 

The occurrence of numerical instabilities in PIC usually depends not only on the numerical parameters like the grid size and time step size, but also on the detailed particle phase space distribution through each step in the PIC computation loop. For example, a PIC simulation with a cold particle distribution can exhibit numerical instabilities with different characteristics than those in a simulation with a Maxwellian particle distribution.  Clearly, for the stability analysis of such a ``numerical plasma", an accurate numerical analysis based on the actual PIC model, not some approximate system, is desirable.   


Similar to the analysis of other dynamical systems, one way to study the numerical stability of the PIC model is to investigate the intrinsic modes the system supports. This is done by assuming a single monochromatic perturbation (usually in the form of a plane wave for a system with translation symmetry) exists on some parts/quantities in the system and then investigating its impact on other parts/quantities of the system and finally the feedback onto the perturbation itself. It is worth noting that the perturbation should be on the internal variables of the isolated system under investigation, not on some external variables that are connected to the corresponding internal variables, e.g., applying an external field on the system or sampling a continuous perturbation into the discrete variable of the system. Furthermore, the perturbation is considered to be smooth without discrete particle noise (i.e., in the limit the number of particle per cell $N \rightarrow \infty$) and to have been turned on for infinite time, so in principle, this is not an initial value problem.    

Mathematically, the above analysis results in an equation describing how a perturbation on some part of the system propagates to other parts of the system and eventually feeds back to itself. For a nonlinear system, this is typically very complicated and one often linearizes the response of each part of the system so as to study the linear properties of the system. We note that this linear approximation, by assuming the perturbation is small, is necessary for the study of the eigen mode of the linear system and should not be confused with other approximations made in order to simplify the algebra, which results in an approximation to the original system. The latter approximations should be avoided as much as possible in the analysis of the numerical model.  Under the linear approximation, the amplitude of the perturbation can be cancelled from the equation. Therefore, one obtains a dispersion relation that describes the intrinsic linear properties of the system. For the ``numerical plasma" model in a PIC code, we will follow the guidelines established above and present a rigorous derivation of the full three dimensional numerical dispersion relation for the standard E.M. PIC algorithm in this paper.  

Our work is motivated by recent interest in the so-called "boosted frame" simulation technique \cite{Vay2007, Martins2010} where the simulation is conducted in a drifting cold plasma background and numerical instability has been observed to distort the results. Although the numerical Cherenkov instability \cite{GODFREY1974, Godfrey1975} is generally believed to be responsible for the unstable modes in the "boosted frame" simulation \cite{Godfrey2013, Xu20132503}, a more fundamental type of numerical instability --- finite grid instability, may be also of interest for the understanding of this technique and other similar simulation models. In particular, the finite grid instability was first analyzed in Ref. \cite{Birdsall-Langdon} for the electrostatic PIC code and a similar analysis for the E.M. PIC code is lacking. Thus our work serves as a bridge between the numerical analysis of the E.S. and E.M. PIC models, as well as a framework for further analysis of the numerical instabilities in a E.M. PIC code. Our reinvestigation of the finite grid instability shows that the above guidelines we seek to apply in this study may not have been fully appreciated in the past. As a result, we find that (1) the temporal aliasing effect should be absent in PIC numerical dispersion; (2) the procedure in which the PIC algorithm advances the particles' phase-space distribution function can be exactly accounted for in the linear stability analysis. The effects of these new findings will be explored in detail in this paper. 

This paper is organized as follows: in Section ~\ref{sec:alias}, we first review how aliasing arises in a discrete system and its implication to the PIC algorithm, then in Section ~\ref{sec:PIC} we describe the standard E.M. PIC algorithm. In Section ~\ref{sec:derivation}, we derive in detail the 3D numerical dispersion of the standard E.M. PIC algorithm. After obtaining the accurate numerical dispersion, we investigate the finite grid instability in E.M. PIC code for the 1D case in Section \ref{sec:finitegrid}. In this section, we focus on the detailed understanding of how such an instability occurs and its dependence on the choice of the numerical parameters. Finally we summarize in section ~\ref{sec:summary}.  

\section{Aliasing}\label{sec:alias}

A major difference between a continuous system and a discrete system is their Fourier space representations. If the system under consideration is inherently discrete (e.g. a crystal lattice), then all information about the system is contained in a single Brillouin zone, with all other Brillouin zones being its exact replicas. If, however, a discretization is imposed on a continuous system, with certain sampling frequency to obtain the discrete system, then every Brillouin zone must be considered to recover a complete description of the discretized system. This aliasing effect (which is defined here in a general sense without considering the bandwidth of the continuous system relative to the Nyquist sampling frequency) is a major source of numerical instability in PIC and also a major complication in the instability analysis. In addition, it is worth noting that the aliasing effect only appears when one continuous system is sampled into a discrete one, not the other way around. Hence sampling is the only cause of aliasing in a numerical model like PIC. 

In fact, both continuous and discrete variables are used in the PIC model, so all three scenarios described above are present. When particle quantities are deposited onto the spatial grid, an effective spatial sampling (but no temporal sampling) is done as the particle distribution function and its derived information, such as the charge density and the current density, are actually known everywhere in space. However, a PIC code implementation does not need to generate the full information and can skip this step. An extra complication exists due to the staggering of particles' positions and velocities in time, so that one can define various particle distribution functions depending on which time steps the position and velocity information are defined on. A particular current deposition scheme may use one of the possible definitions of the particle distribution function. However, this is not the same as knowing the particle distribution function in a continuous time variable. When the field quantities on the grids are interpolated back to the particles' positions, a convolution process instead of a sampling is used. In Fourier space, this is equivalent to the multiplication of the Fourier contents of the fields and the convolution kernel, i.e. the Fourier transform of the particle form factor. For the temporal variable, one can readily recognize that the PIC model operates entirely in a discretized time, without the need to convert back and forth between the continuous and discrete time variables in each computation loop. 

%


Since the only sampling in the PIC algorithm occurs in the spatial variable during the charge and/or current deposition step, when Fourier analyzing a PIC plasma, one should consider all spatial Brillouin zones. No sampling is done to the temporal variable, thus, in Fourier space, solutions to the PIC dispersion relation will be identical in each temporal Brillouin zone. Therefore, all physics of the system can be understood by considering only the zeroth temporal Brillouin zone, and it is sufficient to consider only the contribution from this zone when constructing the Fourier transformed current.

\section{The Electromagnetic PIC Algorithm}\label{sec:PIC}

In this section, we briefly introduce the PIC model to be analyzed --- the standard E.M. PIC algorithm. In this E.M. PIC algorithm, the Yee spatial grid is used  for the fields. 
We will use $\vec{r}_l = \overleftrightarrow{l}\cdot\Delta\vec{r}$, with
\be
\overleftrightarrow{l}=
\begin{pmatrix}
l_x & 0 & 0\\
0 & l_y & 0\\ 
0 & 0 & l_z
\end{pmatrix}
\text{ and }
\Delta\vec{r} =
\begin{pmatrix}
\Delta x\\
\Delta y \\ 
\Delta z
\end{pmatrix},
\nonumber
\ee
 and $l_x, l_y, l_z$ being integers, to denote the position of a grid point and $\vec{r}\,'$ to denote the continuous spatial variable.

\begin{figure}[h!]
  \centering
    \includegraphics[width=0.7\textwidth]{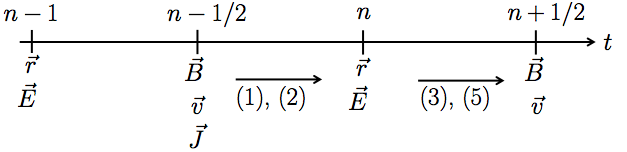}
    \caption{ A schematic of how the system is advanced in an E.M. PIC model. Suppose $\vec{r}_{n-1}$, $\vec{E}_{n-1}$, $\vec{v}_{n-1/2}$ and $\vec{B}_{n-1/2}$ are known. First, a half-time-step push is carried out using Eq. (\ref{eq:acc}) to get $\vec{r}_{n-1/2}$, which is used with $\vec{v}_{n-1/2}$ to obtain the particle distribution function at time step $(n-1/2)$ and then $\vec{J}_{n-1/2}$. Another half push is performed to get $\vec{r}_n$, then Eq. (\ref{eq:amp}) is used to get $\vec{E}_n$. Eqs. (\ref{eq:far}) and (\ref{eq:bor}) are then used to obtain $\vec{v}_{n+1/2}$ and $\vec{B}_{n+1/2}$.}
\label{fig:time}
\end{figure}

The equation of motion and the Maxwell's equations are discretized with central differencing. Both the particles and the electromagnetic fields are advanced with the leap-frog scheme. In particular, particle position $\vec{r}$, velocity $\vec{v}$, and the fields $\vec{E}, \vec{B}$ are advanced in time according to

\begin{align}
\frac{d\vec{r}}{dt}=\vec{v} &\longrightarrow \frac{\vec{r}_{n-1/2}-\vec{r}_
{n-1}}{\Delta t/2} = \vec{v}_{n-1/2} \text{ and }
\frac{\vec{r}_{n}-\vec{r}_{n-1/2}}{\Delta t/2} = \vec{v}_{n-1/2}\label{eq:acc}
\\
\frac{\partial\vec{E}}{\partial t} = \vec{\nabla}\times\vec{B} - 4\pi\vec{J} &\longrightarrow \frac{\vec{E}_n-\vec{E}_{n-1}}{\Delta t} = \vec{\nabla}\times\vec{B}_{n-1/2}-4\pi\vec{J}_{n-1/2}\label{eq:amp}
\\
\frac{\partial\vec{B}}{\partial t} = -\vec{\nabla}\times\vec{E} &\longrightarrow \frac{\vec{B}_{n+1/2}-\vec{B}_{n-1/2}}{\Delta t} = -\vec{\nabla}\times\vec{E}_n\label{eq:far}
\\
\frac{d\left(\gamma\vec{v}\right)}{d t} = \frac{q}{m}\left(\vec{E} + \vec{v}\times\vec{B}\right) &\longrightarrow \frac{\left(\gamma\vec{v}\right)_{n+1/2}-\left(\gamma\vec{v}\right)_{n-1/2}}{\Delta t} = \frac{q}{m}\left(\vec{E}_n + \vec{v}_n\times\vec{B}_n\right), \label{eq:lor}
\end{align}
where $\gamma= 1/\sqrt{1-(v/c)^2}, q, m$ are the particle Lorentz factor, charge and mass, respectively. The equations for the continuous system is shown on the left and their discrete analogs used in the PIC algorithm is shown on the right. Here, $n$ is the index of time step, $t = n\Delta t$.  With the Boris algorithm for particle pushing, Eq. (\ref{eq:lor}) takes on the form of a pure rotation,
\be
\frac{\left(\gamma \vec{v} \right)^+-\left(\gamma \vec{v} \right)^-}{\Delta t} = \frac{q}{2m\gamma_n}\left[ \left(\gamma \vec{v} \right)^++\left(\gamma \vec{v} \right)^-\right] \times\vec{B}_n
\label{eq:bor}
\ee
and two half accelerations $\left(\gamma\vec{v}\right)_{n\pm 1/2} = \left(\gamma \vec{v} \right)^\pm \pm q\vec{E}_n\Delta t/2m$. Eqs (\ref{eq:acc})-(\ref{eq:bor}) are implemented as shown in Fig. \ref{fig:time}.

In Eq. \ref{eq:amp}, the current density is calculated at the grid points where $\vec{E}$ is defined.  \emph{This is exactly equivalent to spatial sampling}, as the underlying particle distribution, from which the current density is obtained, is defined on the continuous spatial variable. There are many current deposition schemes developed for the E.M. PIC code, among those the charge conserving schemes in \cite{Villasenor1992} and \cite{Esirkepov2001} are widely-used. Although the exact details of these schemes differ, one can attribute their differences being the particle distribution function used in the deposition step. For example, some current deposition schemes use the concept of` ``virtual particle" for the deposition step. The ``virtual particles" are created through a transform on the actual particle distribution at one time step or actual distributions on several consecutive time steps (but not at arbitrary continuous time so this is not a form of sampling in time). Similarly, other current deposition schemes define their own transforms. In this paper, we ignore these differences in the transform and analyze the standard E.M. PIC with a simple current deposition scheme where the particle distribution function is used directly to obtain the current density at the same step. This is shown in Fig. \ref{fig:time}, where the particle distribution at $t_{n-1/2}$ is first obtained through a half position update and then the current density is deposited. For this simple scheme, the current density at the grid points, $\vec{J}\left(\vec{r}_l,t_{n-1/2}\right)$, is related to the current density in the continuous space, $\vec{J}\left(\vec{r}\,',t_{n-1/2}\right)$, according to
\be
\vec{J}\left(\vec{r}_l,t_{n-1/2}\right) = \int \overleftrightarrow{S_J}\left(\vec{r}\,'-\vec{r}_l\right)\cdot\vec{J}\left(\vec{r}\,',t_{n-1/2}\right)d^3\vec{r}\,'.
\label{eq:Jdisc}
\ee
Here, $\overleftrightarrow{S_J}\left(\vec{r}\,'-\vec{r}_l\right)$ is the current interpolation tensor which is determined by the particle shape. For a more complicated current deposition scheme, the particle distribution transform will introduce an additional factor in the integral. A similar convolution expression is used to interpolate the fields at the grid points to the particle's position (which will be given in the next section). However, no aliasing will result as no sampling is involved.   

\section{Derivation of the numerical dispersion of electromagnetic PIC code}\label{sec:derivation}

In this section, we will derive the numerical dispersion step by step.

\subsection{Finite Difference Wave Equation and Current Deposition in Fourier Space}

To obtain the finite-difference dispersion relation for the PIC algorithm, we'll use the normalized quantities for the electric field $\vec{E} \rightarrow e \vec{E} / mc\omega_{p}$, magnetic field $\vec{B} \rightarrow e \vec{B} / mc\omega_{p}$, temporal coordinate $t \rightarrow \omega_{p}t$, spatial coordinate $\vec{r} \rightarrow \omega_{p}\vec{r}/c$, velocity $\vec{v} \rightarrow \vec{v}/c$, temporal Fourier component $\omega \rightarrow \omega/\omega_p$, spatial Fourier component  $\vec{k} \rightarrow \vec{k}/k_p$, momentum $\vec{p} \rightarrow \vec{p}/mc$, electric charge $q \rightarrow q/e$, number density $n \rightarrow n/n_{p}$, electric current density $\vec{J} \rightarrow \vec{J}/en_{p}c$. Here, $m$ is the electron mass, $e$ is elementary charge, $c$ is luminal vacuum speed, $n_p$ is the electron plasma density, and $\omega_p$ is the electron plasma frequency. We begin with the Fourier transformed Amp$\grave{\text{e}}$re and Faraday laws,
\begin{eqnarray}
i\vec{k}_B\times\tvB=\tvJ-i\omega_E\tvE \nonumber \\
\vec{k}_E\times\tvE=\omega_B\tvB,
\label{eq:ftM}
\end{eqnarray}
where
\be
\begin{array}{r@{}l}
&{} \vec{E}\left(\vec{r}_l,t_{n-1}\right)=\tvE(\vec{k},\omega)e^{i\left(\vec{k}\cdot\vec{r}_l - \omega t_{n-1} \right)}  \cr
&{} \vec{B}\left(\vec{r}_l,t_{n-1/2}\right)=\tvB(\vec{k},\omega)e^{i\left(\vec{k}\cdot\vec{r}_l - \omega t_{n-1/2}\right)} \\
&{} \vec{J}\left(\vec{r}_l,t_{n-1/2}\right)=\tvJ(\vec{k},\omega)e^{i\left(\vec{k}\cdot\vec{r}_l - \omega t_{n-1/2}\right)}.  \cr
\label{eq:EBJFourier}
\end{array}
\ee
In Eqs. (\ref{eq:ftM}),  $\tvB$, $\tvE$, and $\tvJ$ are Fourier transformed fields and current density, $\vec{k}_E$, $\vec{k}_B$, and $\omega_E$, $\omega_B$ are the Fourier transformed spatial and temporal finite difference operators for the fields. Combining Eqs. (\ref{eq:ftM}) yields the Fourier transformed finite difference wave equation,
\be
\vec{k}_B\times\vec{k}_E\times\tvE+\omega _B\omega _E\tvE=- i\omega _B\tvJ.
\label{eq:weqn}
\ee

To evaluate Eq. (\ref{eq:weqn}), we need an expression for the Fourier transformed current, $\tvJ\left(\vec{k},\omega\right)$. Let us assume a linearized kinetic model for a uniform, charge-neural and current-free plasma consisting of drifting cold electrons of charge $q_e$ and rest mass $m$, and co-moving neutralizing ions with infinite mass. For this model, we only consider the electron response. The distribution function for the electrons is
\be
f\left(\vec{r}\,',\vec{p}, t \right)=f_0\left(\vec{p}\right)+f_1\left(\vec{r}\,',\vec{p}, t \right),
\label{eq:dist}
\ee
with $\left|f_1/f_0\right|<<1$ and $\int f_0(\vec{p}) d \vec{p} = 1$.

At time $t_{n-1/2}=\left(n-1/2\right)\Delta t$, the current in continuous real space is 
\be
\vec{J}\left(\vec{r}\,',t_{n-1/2}\right)=q_e \int\frac{\vec{p}}{\gamma }f_1\left(\vec{r}\,',\vec{p}, t_{n-1/2} \right)d^3\vec{p}.
\label{eq:Jcont}
\ee

Combining Eqs. (\ref{eq:Jdisc}) and (\ref{eq:Jcont}), it follows that
\be
\vec{J}\left(\vec{r}_l,t_{n-1/2}\right) = q_e \int\int \overleftrightarrow{S_J} \left(\vec{r}\,'-\vec{r}_l\right)\cdot\frac{\vec{p}}{\gamma} f_1\left(\vec{r}\,', \vec{p}, t_{n-1/2} \right)d\vec{p}\,d\vec{r}\,'.
\label{eq:Jx}
\ee

This is the expression for the current deposition step in the PIC algorithm where a spatial sampling is carried out implicitly. No temporal sampling is done as both sides of the equation are defined on the same discrete time. For the linear dispersion analysis presented here, the fully discretized current $\vec{J}\left(\vec{r}_l,t_{n-1/2}\right) $ can be written in the form of a monochromatic plane wave as in Eq. (\ref{eq:EBJFourier}).

Eq. (\ref{eq:Jx}) also indicates that $f_1$ has the same dependence on $\omega$ as $\vec{J}\left(\vec{r}_l,t_{n-1/2}\right) $. On the other hand, one cannot assume that they have the same dependence in $\vec{k}$, due to aliasing resulting from spatial sampling. However, since $\vec{r}\,'$ is continuous, we can write 
\be
f_{1}(\vec{r}\,',\vec{p},t_{n-1/2})=e^{-i\omega t_{n-1/2}}\int\tilde{f}_{1}(\vec{k},\vec{v},\omega)e^{i\vec{k}\cdot\vec{r}\,'}d\vec{k} \text{ and }
\overleftrightarrow{S_J} (\vec{r}\,')=\int\tilde{\overleftrightarrow{S_J}}(\vec{k})e^{i\vec{k}\cdot\vec{r}\,'}d\vec{k}.
\label{eq:ftf}
\ee

From Eqs. (\ref{eq:Jx}) - (\ref{eq:ftf}), we obtain 
\begin{equation}
\tvJ(\vec{k}, \omega)e^{i\vec{k}\cdot\vec{r}_{l}}= q_e \int\tilde{\overleftrightarrow{S_J}}(\vec{k}')\left[\int\frac{\vec{p}}{\gamma}\tilde{f}_{1}(\vec{k}',\vec{p},\omega)d\vec{p}\right]e^{i\vec{k}'\cdot\vec{r}_l}d\vec{k}'.
\end{equation}

Since $\tvJ \neq 0$ only at the grid points in the reciprocal space,
\begin{equation}
\tilde{\overleftrightarrow{S_J}}(\vec{k}')\left[\int\vec{p}\tilde{f}_{1}(\vec{k}',\vec{p},\omega)d\vec{p}\right]\propto\delta(\vec{k}'-\vec{k}-\vec{q}),
\end{equation}
with $\vec{q} = 2\pi\left(q_x/\Delta x,q_y/\Delta y,q_z/\Delta z\right)$ being a vector between grid points in the reciprocal 3-D lattice. As a direct result
\begin{equation}
\tvJ(\vec{k}, \omega)= q_e \sum_{\vec{q}}\tilde{\overleftrightarrow{S_J}}(\vec{k}_q)\left[\int\frac{\vec{p}}{\gamma}\tilde{f}_{1}(\vec{k}_q,\vec{p},\omega)d\vec{p}\right],
\label{eq:ftJ}
\end{equation}
where $\vec{k}_q = \vec{k}+\vec{q}$.
The wave equation Eq. (\ref{eq:weqn}) now takes on the form
\be
\vec{k}_B\times\vec{k}_E\times\tvE+\omega _B\omega _E\tvE=-i q_e \omega _B\sum_{\vec{q}}\tilde{\overleftrightarrow{S_J}}(\vec{k}_q)\left[\int\frac{\vec{p}}{\gamma}\tilde{f}_{1}(\vec{k}_q,\vec{p},\omega)d\vec{p}\right],
\ee
where aliasing has been taken into account in the infinite sum. We obtain an expression for $\tilde{f}_1$ in the next section.

\subsection{Expression for the $1^{st}$ order distribution function, $\tilde{f}_{1}$}

In a continuous system, the first order distribution function $f_1 \left(\vec{r}\,',\vec{p}, t \right)$ in Eq. (\ref{eq:dist}) satisfies the linearized Vlasov equation
\be
\frac{\partial f_1}{\partial t}+\frac{\vec{p}}{\gamma}\cdot\vec{\nabla} f_{1}+ \vec{F} \cdot\vec{\nabla}_pf_0 =0.
\label{eq:vlas}
\ee
where $\vec{F}=q_e \left[\vec{E}(\vec{r}\,',t)+(\vec{p}/ \gamma) \times\vec{B}(\vec{r}\,',t)\right]$ is the Lorentz force.

However, in the PIC algorithm, Eq. (\ref{eq:vlas}) is solved in three steps: beginning from a particle distribution with position and velocity known at time step $(n-1/2)$, i.e., after current deposition, first step is a half-time-step position push, followed by a full-time-step momentum push, then another half-time-step position push. After these three steps, the particle distribution is advanced to time step $(n+1/2)$. The first half position push is equivalent to solving 
\be
\frac{\partial f_1}{\partial t} + \frac{\vec{p}}{\gamma} \cdot \vec{\nabla} f_1 = 0 
\label{eq:half-push}
\ee
for a half-time-step. It is important to note that this is accomplished in PIC using the method of characteristics instead of Fourier transforming or using a finite difference approximation of Eq. (\ref{eq:half-push}). If the first order distribution function used for the current deposition at time step $(n-1/2)$ is taken to be $f_1\left(\vec{r}\,', \vec{p}, t_{n-1/2} \right) = \tilde{f}_{1} \left( \vec{k}, \vec{p}, \omega \right) e^{i\left(\vec{k}\cdot\vec{r}\,'-\omega t_{n-1/2} \right)}$, then after the half position push, the updated distribution function $f_1^{*}\left( \vec{r}\,', \vec{p} \right) = \tilde{f}_1^{*}  \left( \vec{k}, \vec{p}, \omega \right)  \allowbreak e^{i\left(\vec{k}\cdot\vec{r}\,'-\omega t_{n-1/2} \right)}$ is   
\be
f_1^{*}\left( \vec{r}\,', \vec{p} \right)=f_1\left(\vec{r}\,'-\vec{v}\Delta t/2,  \vec{p}, t_{n-1/2}\right),
\label{eq:fs}
\ee
where the time argument of $f_1^{*}$ has been suppressed for clarity as this distribution function is constructed using the particle position and velocity at different time steps. It follows that, 
\be
\tilde{f}^*_1\left( \vec{k}, \vec{p}, \omega \right)=\tilde{f}_1\left( \vec{k}, \vec{p}, \omega \right) e^{-i \vec{k}\cdot\vec{v} \Delta t/2}.
\label{eq:first-push}
\ee

Next we look at the full momentum push, which is equivalent to solving 
\be
\frac{\partial f_1}{\partial t} + \vec{F}\cdot\vec{\nabla}_p f_0\left(\vec{p}\right)=0.
\ee
This is accomplished in PIC using the Boris pusher and the updated distribution $f_{1}^{**}\left(\vec{r}\,',  \vec{p} \right)$ can be written as,
\be
f_{1}^{**}\left(\vec{r}\,',  \vec{p} \right) =\int \left[f_0\left(\vec{p}\,'  \right)+f_{1}^{*}\left(\vec{r}\,',  \vec{p}' \right) \right] \delta \left(\vec{p} - \vec{p}\,'  - \Delta \vec{p} \right) d\vec{p}\,' - f_0\left(\vec{p}\right),
\ee
with $\Delta\vec{p}=\vec{F}\Delta t$ \footnote{Note that $\vec{p}$ is the momentum axis coordinate and is continuous. It is \emph{not} the individual particles' momenta and is \emph{not} a function of the discrete time coordinate.}. 

Taylor expanding the $\delta$-function about $\vec{p}=\vec{p}\,'$, linearizing, then integrating by parts gives
\be
f_{1}^{**}\left(\vec{r}\,',  \vec{p} \right) \approx - \int  \Delta \vec{p} \cdot \vec{\nabla}_{p'} f_0\left(\vec{p}\,' \right) \delta\left(\vec{p}-\vec{p}\,' \right)  d \vec{p}\,' + \int f^*_1\left(\vec{r}\,',  \vec{p}\,'  \right) \delta\left(\vec{p} -\vec{p}\,' \right) d\vec{p}\,'
\ee
or,
\be
f_{1}^{**}\left(\vec{r}\,',  \vec{p} \right)  \approx -\vec{\nabla}_{p} f_0\left(\vec{p} \right) \cdot \Delta\vec{p}+f^*_1 \left(\vec{r}\,',  \vec{p} \right).
\ee
Note that $\vec{F}$ is equal to the sum of two terms -- one involving the perturbation $\vec{E}_1$, and one involving the perturbation $\vec{B}_1$. Both terms are expressed as monochromatic plane waves, then we must have $\vec{F}\left(\vec{r}\,',  \vec{p}, t_{n}\right)=\tilde{\vec{F}}(\vec{k},\vec{p},\omega) e^{i\left(\vec{k}\cdot\vec{r}\,'-\omega t_{n-1/2} \right)}$. Note that we use the same time harmonic form $e^{i\omega t_{n-1/2}}$ as in $\vec{J}( \vec{r}_l, t_{n-1/2})$ for $\vec{F}(t_n)$ here and leave the discussion of the spatial and temporal phase factors relating $\vec{E}$, $\vec{B}$ and $\vec{F}$ to the next section. Therefore,
\be
\tilde{f}^{**} _1\left(\vec{k},\vec{p},\omega\right)=-\Delta t \tilde{\vec{F}}\left(\vec{k},\vec{p},\omega\right)\cdot\vec{\nabla}_p f_0\left(\vec{p}\right) + \tilde{f}_1 \left(\vec{k},\vec{p},\omega\right) e^{-i \vec{k}\cdot\vec{v} \Delta t/2}.
\label{eq:momentum-update}
\ee

For the second half position push, we solve for the updated distribution function like we did for the first half push. The updated distribution $f_{1}^{***}\left(\vec{r}\,',\vec{p},t_n\right)$ becomes, 

\be
f_{1}^{***}\left(\vec{r}\,',\vec{p},t_n\right) = f_{1}^{**} \left(\vec{r}\,' - \vec{v}\Delta t/2, \vec{p},t_n\right).
\label{eq:second-push}
\ee

Combining Eqs. (\ref{eq:first-push}), (\ref{eq:momentum-update}) and (\ref{eq:second-push}), we obtain 
\be
\tilde{f}^{***}_1\left(\vec{k},\vec{p},\omega\right)= - \Delta t \tilde{\vec{F}}\left(\vec{k},\vec{p},\omega\right) \cdot \vec{\nabla}_p Ÿf_0 e^{-i \vec{k} \cdot \vec{v} \Delta t/2}+\tilde{f}_1\left(\vec{k},\vec{p},\omega\right) e^{-i\vec{k}\cdot\vec{v}\Delta t}.
\ee
Since one full time-step has elapsed after these three steps, $\tilde{f}^{***}_1\left(\vec{k},\vec{p},\omega\right)=\tilde{f}_1 \left(\vec{k},\vec{p},\omega\right)e^{-i\omega\Delta t}$.  Therefore,
\be
\tilde{f}_{1}(\vec{k}, \vec{p}, \omega)=\frac{ \tilde{ \vec{F} } \left(\vec{k}, \vec{p}, \omega \right) \cdot\vec{\nabla}_p f_{0} e^{-i \vec{k}\cdot\vec{v} \Delta t/2}} {(e^{-i\vec{k}\cdot\vec{v}\Delta t}-e^{-i\omega\Delta t})/\Delta t}.
\label{eq:f1t}
\ee

Eq. (\ref{eq:f1t}) includes the effect of a finite time step on $\tilde{f}_1$. When $\Delta t \rightarrow 0$, $\tilde{f}_1 \rightarrow \tilde{\vec{F}}\cdot\vec{\nabla}_p f_0/ \left(\omega - \vec{k}\cdot \vec{v}\right)$, which recovers  the expression for a system with continuous time variable. Now we are in need of yet another expression: $\tilde{\vec{F}}$. We find this  in the following section by paying careful attention to the spatial and temporal phase factors.

\subsection{Expression for the Fourier transformed Lorentz force, $\tilde{\vec{F}}$}

Before we proceed, it is important to ensure that all the terms in each equation are evaluated at the same time step with the same spatial grid. With the use of a Yee grid for the fields, Eq.  (\ref{eq:ftM}) implies that each spatial component of these equations are defined at the same grid points (i.e., the reference phase on this grid has been factored out) as the field not involved in the spatial derivative (e.g., the $x$ component of Faraday's law is defined at the same grid points as $B_x$, and the $x$ component of Amp$\grave{\text{e}}$re's law is defined at the same grid points as $E_x$, etc.) It is conventional to define $\omega _B = \omega \text{sinc}\left(\omega \Delta t/2\right)$ and $\vec{k}_E = \left[k_x \text{sinc}\left(k_x \Delta x/2\right),k_y \text{sinc}\left(k_y \Delta y/2\right),k_z \text{sinc}\left(k_z \Delta z/2\right)\right]$ in Faraday's law, because the standard PIC algorithm uses a centered scheme when calculating derivatives.\footnote{For non-standard PIC algorithms, such as \cite{Greenwood2004}, the forms of these operators are different thus affecting the numerical dispersion and instability. The numerical dispersion analysis of these algorithms will be presented in a future publication.} Similarly, for Amp$\grave{\text{e}}$re's law in Eq. (\ref{eq:ftM}), we choose $\omega_E = \omega_B$ and $\vec{k}_B = \vec{k}_E$. We will explicitly account for any implicit phase factors that may be present due to the staggered temporal and spatial grids, the choice of the reference phase, and the form of the finite difference operators.

In Amp$\grave{\text{e}}$re's law from Eq. (\ref{eq:ftM}), $\tilde{\vec{J}}$ is calculated using $\vec{J}$ evaluated at the time step of $\vec{B}$ and the spatial grid of $\vec{E}$. The expression for $\vec{J}$ is given by Eq. (\ref{eq:Jx}), and is proportional to $f_1$ at the same time step. Furthermore, $f_1$ is proportional to the Lorentz force, involving $\vec{E}$ and $\vec{B}$, for which the same reference phase for the time step and grid should be used. However, the leap-frog advance requires the Lorentz force to be evaluated a half-time-step ahead of $f_1$ and $\vec{B}$. $\vec{E}$ is defined on this time step so it does not require time averaging, but a temporal phase factor is needed to account for the half-time-step difference relative to $f_1$. $\vec{B}$ is not defined on this time step and thus requires time averaging. In order to account for the different spatial phases factored out from Eq. (\ref{eq:ftM}) and the Lorentz force, one also needs to consider the spatial phase difference between the staggered spatial grids relative to that of $\vec{E}$. The expression for the Lorentz force, which is extrapolated to the particle's position $\vec{r}$ and includes all the phase factors mentioned above, is given by

\begin{align}
 \vec{F}\left(\vec{r},t_n\right) & =  \tilde{\vec{F}} e^{i(\vec{k} \cdot \vec{r} - \omega t_{n-1/2})}    \notag \\
  & =   q_e  \sum_{\vec{r}_{l}} e^{i(\vec{k} \cdot \vec{r}_l - \omega t_{n-1/2})} \left[ \tau_E \overleftrightarrow{S_E} \left(\vec{r}-\vec{r}_{l} \right) \cdot \overleftrightarrow{O_E} \cdot  \tilde{\vec{E}}  +\tau_B (\overleftrightarrow{O_B} \cdot \frac{\vec{p}}{\gamma} ) \times \overleftrightarrow{S_B} \left(\vec{r}-\vec{r}_l\right) \cdot \tilde{\vec{B}}  \right] , \notag \\
\label{eq:lorentz} 
\end{align}
where the temporal phase factors are $\tau_E=e^{-i\omega\Delta t/2}$ and $\tau_B=\left(e^{-i\omega \Delta t}+1\right)/2$, which results from time averaging. The spatial phase tensors are $\overleftrightarrow{O_E}$ and  $\overleftrightarrow{O_B}$, and $\vec{r}_l$ is the spatial grid of $\vec{E}$ for the corresponding component of the Lorentz force.

Now we need the elements of the spatial phase tensors $\overleftrightarrow{O_E}$ and $\overleftrightarrow{O_B}$. Because the spatial grid of $\vec{E}$ is the reference grid, it is clear that  $\overleftrightarrow{O_E} = \overleftrightarrow{I_3}$, i.e., the identity matrix. We also need the spatial phase in every component of the magnetic force to be the same as the spatial phase of the electric part. That is, the phase relationships are enforced among,
\begin{eqnarray}
E_x &&\leftrightarrow v_y O_B^{xz} B_z \text{ and } v_z O_B^{xy} B_y, \nonumber \\
E_y &&\leftrightarrow v_x O_B^{yz} B_z \text{ and } v_z O_B^{yx} B_x, \nonumber \\
E_z &&\leftrightarrow v_y O_B^{zx} B_x \text{ and } v_x O_B^{zy} B_y. \nonumber
\end{eqnarray}

Because of the way the components of $\vec{E}$ and $\vec{B}$ are staggered on the Yee grid, we can obtain, 
\begin{eqnarray}
O_B^{xz} = e^{ik_y\Delta y/2} && O_B^{xy} = e^{ik_z\Delta z/2} \nonumber \\
O_B^{yz} = e^{ik_x\Delta x/2} && O_B^{yx} = e^{ik_z\Delta z/2} \nonumber \\
O_B^{zx} = e^{ik_y\Delta y/2} && O_B^{zy} = e^{ik_x\Delta x/2}, \nonumber
\end{eqnarray}
where $O_B^{ij}$ is a scalar phase factor.
This can be expressed compactly as
\be
\overleftrightarrow{O_B}=
\begin{pmatrix}
e^{ik_x\Delta x/2} & 0 & 0\\
0 & e^{ik_y\Delta y/2} & 0\\ 
0 & 0& e^{ik_z\Delta z/2}
\end{pmatrix}
\ee
which is used in Eq.  (\ref{eq:lorentz}).

From Eqs. (\ref{eq:ftM}) and (\ref{eq:lorentz}), 
\be
\tilde{\vec{F}}\left(\vec{k}, \vec{p}, \omega\right) = q_e \left[ \tau_E \tilde{\overleftrightarrow{S_E}} \left(\vec{k}\right)\cdot\tvE\left(\vec{k},\omega\right) + \frac{\tau_B}{\omega_B} \left(  \overleftrightarrow{O_B}  \cdot \frac{\vec{p}}{\gamma } \right) \times\tilde{\overleftrightarrow{S_B}}\left(\vec{k}\right) \cdot \vec{k}_E \times \tvE\left(\vec{k},\omega\right)  \right]
\ee

We finally have an explicit expression for the current, 
\begin{equation}
\tvJ(\vec{k}, \omega)=\sum_{\vec{q}}\tilde{S}_{j}(\vec{k}_q)\left[\int\vec{p}\frac{\tilde{\vec{F}}\left(\vec{k}_q, \vec{p}, \omega\right) \cdot\vec{\nabla}_p f_{0}e^{-i\left(\vec{k}_q\cdot\vec{v}-\omega\right)\Delta t/2}}{\gamma(e^{-i\vec{k}_q\cdot\vec{v}\Delta t}-e^{-i\omega\Delta t})/\Delta t}d\vec{p}\right].
\label{eq:ftJ}
\end{equation}
Substituting this into the Fourier transformed wave equation, Eq. (\ref{eq:weqn}), eliminating $\tvB$ from $\tvJ$ using Faraday's law from Eq. (\ref{eq:ftM}), and noticing that $\tvE (\vec{k}_q, \omega) = \tvE (\vec{k}, \omega)$, we obtain an equation of the form
\be
\overleftrightarrow{\epsilon}\cdot\tvE (\vec{k}, \omega)= 0.
\ee
This has nontrivial solutions for $\tvE(\vec{k}, \omega)$ only if

\be
\text{Det}\left|\overleftrightarrow{\epsilon}\right|=0
\label{eq:disprel}
\ee
Eq. (\ref{eq:disprel}) is the finite difference dispersion relation relating $\vec{k}$ and $\omega$.

\section {Numerical Instability from Finite Difference Modes for a Relativistically Drifting Plasma}\label{sec:finitegrid}

Let us consider the case where the unperturbed distribution function is that for a cold relativistically drifting plasma. For the simplest case, the drift can be taken to be in the $\hat{x}$ direction with each electron having momentum of magnitude $p_0$. Then,
\be
f_0\left(\vec{p}\right)=\delta\left(p_x-p_0\right)\delta\left(p_y\right)\delta\left(p_z\right).
\ee

In the case of a 1-D Yee algorithm, the dispersion relation Eq. (\ref{eq:disprel}) becomes
\be
\omega_B
\begin{vmatrix}
\epsilon _a & 0 & 0\\
0 & \epsilon _b & 0\\ 
0 & 0 & \epsilon _b
\end{vmatrix} = 0,
\label{eq:disp}
\ee
where
\be
\begin{split}
&\epsilon_a=\frac{\sin \left(\omega  \Delta t/2\right)}{\Delta t /2}+\frac{e^{i\omega\Delta t}\Delta t}{\gamma ^3}\times \\
&\sum_{q=-\infty}^\infty \frac{2 \sin \left\{\left(\omega-k_q v_0 \right)\Delta t/2\right\}+k_q v_0 \Delta t \cos \left\{\left(\omega -k_q v_0\right)\Delta t/2\right\}}{  \left(e^{i \omega  \Delta t}-e^{i k_q v_0 \Delta t}\right){}^2}\text{sinc} ^4\left( k_q \Delta x/2 \right) e^{i k_q v_0 \Delta t},
\end{split}
\label{eq:ESdxdt}
\ee
and
\be
\begin{split}
\epsilon_b=\frac{\sin ^2\left(\omega \Delta t/2\right)}{\left(\Delta t/2\right)^2}-&\frac{\sin ^2\left(k \Delta x/2\right)}{\left(\Delta x/2\right)^2}+\frac{e^{-i \omega  \Delta t/2}}{2 \gamma  \Delta x}\sum_{q =-\infty}^\infty \csc \left\{\left(\omega-k_q v_0 \right)\Delta t/2 \right\}\times\\
& \left\{v_0 \Delta t \left(1+e^{i \omega  \Delta t}\right) e^{i k_q \Delta x/2} \sin \left(k_q \Delta x/2\right)+i \Delta x \left(e^{i \omega  \Delta t}-1\right)\right\},
\end{split}
\ee
and
\be
\omega_B=\frac{\sin \left(\omega  \Delta t/2\right)}{\Delta t /2},
\ee
with $k_q = k+2\pi q/\Delta x$, $v_0 = p_0 / \gamma_0$. Note that the directional subscripts have been omitted since only one direction is being considered.
Eq. (\ref{eq:disp}) is clearly satisfied when any of the following conditions are met for $\omega \neq 0$ ($\omega_B=0$ only gives $\omega = 0$ mode) :
\begin{eqnarray}
\epsilon_a=0&
\label{eq:esDisp}\\
\epsilon_b=0&
\label{eq:emDisp}
\end{eqnarray}


In the limit we take $\Delta x \rightarrow 0$, $\Delta t \rightarrow 0$, and set $k_q=k$, Eqs. (\ref{eq:esDisp}) and (\ref{eq:emDisp}) become the dispersion equations for the Electrostatic-like (E.S.-like) and Electromagnetic-like (E.M.-like) modes of a drifting cold plasma
\footnote{ We add the "-like" suffix, because for $\gamma>1$ the E.S.-like eigenvector is not parallel to the wave vector. Similarly, the second mode is E.M.-like, as its eigenvector is not perpendicular to its wave vector for $\gamma>1$ \cite{McKinstrie1995}.}. They are, respectively, 
\begin{eqnarray}
\omega^2 -k^2  &=& 1/\gamma, \nonumber\\
\left(\omega-k v_0\right)^2 &=& 1/\gamma^3. \nonumber
\end{eqnarray}


With a finite grid and time-step, Eqs. (\ref{eq:esDisp}) and (\ref{eq:emDisp}) are transcendental in $\omega$ and $k$ and must be solved numerically. For n this paper, We will focus on the E.S.-like mode dispersion relation, Eq. (\ref{eq:esDisp}),  . Before we delve into solving the full finite difference dispersion relations, we consider the simpler case with $\Delta t \rightarrow 0$ and only a finite grid size $\Delta x$.

\subsection{Instability from the Electrostatic-like mode for $\Delta t \rightarrow 0$}

It is well known that PIC codes exhibit a type of numerical instability due to the use of a finite size grid. The analysis of this instability can be made using the drifting cold plasma model in the limit of an infinitely small  time step size \cite{Birdsall-Langdon}. Taking this limit as $\lim_{\Delta t \rightarrow 0}\epsilon_a  = 0$, we find that the finite difference dispersion relation for the E.S.-like mode becomes, 
\be
\left(\omega_t+k v_0\right)\left[ 1 - \frac{16 \Delta x^2}{\gamma^3}\sin^4\left({k \Delta x/2}\right)
  \sum_{q=-\infty}^{\infty}\frac{1}{\left(k \Delta x - 2 \pi q\right)^4 \left(\omega_t\Delta x + 2 \pi q v_0\right)^2} \right]=0
\label{eq:dt0}
\ee
where $\omega_t = \omega - k v_0$. The first factor corresponds to a second standalone $\omega=0$ mode and is not of particular interest. The infinite sum is easily evaluated with Cauchy's residue theorem so that the second factor in Eq. (\ref{eq:dt0}) becomes

\be
\begin{split}
&\frac{1}{\left(\omega_t+k v_0\right){}^5}\left\{-\Delta x^2 \left(\omega_t + k v_0\right){}^2 \sin ^2\left(\omega _t \Delta x/2 v_0\right) \left\{\Delta x \left(\omega_t+k v_0\right) \left\{3 \gamma ^3 \left(\omega_t+k v_0\right){}^2\right. \right. \right.\\
&\left. \left. \left. -\cos \left(k \Delta x\right)-2\right\}-6 v_0 \sin \left(k \Delta x\right)\right\}+12 v_0^2 \sin ^4\left(k \Delta x/2\right) \left\{\Delta x \left(\omega_t+k v_0\right) +4 v_0 \sin \left(\omega _t \Delta x / v_0\right)\right\}\right. \\
&\left. +12 v_0^2 \sin ^2\left(k \Delta x/2\right) \sin ^2\left(\omega _t \Delta x/2 v_0\right) \left\{3 \Delta x \left(\omega_t+k v_0\right)+4 v_0 \sin \left(k \Delta x_1\right)\right\}\right\}=0.
\end{split}
\label{eq:dt0exanded}
\ee

\begin{figure}[h!]
  \centering
    \includegraphics[width = \textwidth]{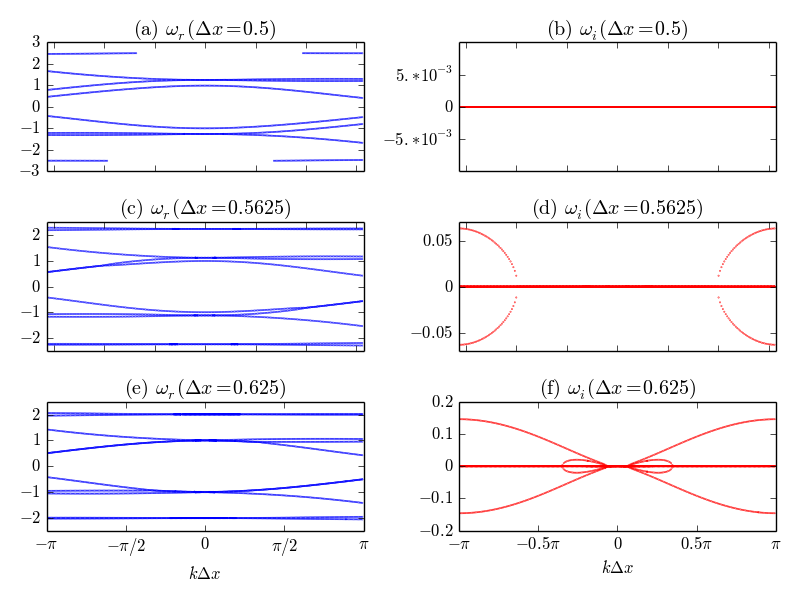}
    \caption{Numerical solutions in the zeroth spatial Brillouin zone $k\Delta x \in \left[-\pi,\pi\right]$ to the dispersion relation Eq. (\ref{eq:dt0}), for $v_0 = 0.1$ and with varying grid spacing $\Delta x$. }
    \label{fig:dt0inst}
\end{figure}

We now set $\omega_t = \omega_r + i \omega_i$, and numerically search for solutions in the complex plane, given a set of numerical parameters $k$, $\Delta x$, and $v_0$. Fig. \ref{fig:dt0inst} shows the solutions for $\omega_r$ and $\omega_i$ as functions of $k\Delta x$ for $v_0=0.1$. In general, the solutions in $\omega_r$ consist of a finite grid size plasma mode and alias modes corresponding to the poles from the sum in the second factor in Eq. (\ref{eq:dt0}). Therefore, they can be labeled by the $q$ index of the pole (the $q=0$ mode is the finite grid size plasma E.S.-like mode, while $q \ne 0$ modes are alias modes). Since the vertical location of the pole is determined by $\omega_r=-2 \pi q v_0/ \Delta x$, overall these modes are parallel to each other and also to the $\omega_r = 0$ axis, and may not cross (we note that similar observation applies to the electrostatic PIC dispersion relation Eq. (7) derived in chapter 8-11 of Ref. \cite{Birdsall-Langdon} for $\Delta t \rightarrow 0$, but our calculation shows that each alias mode from that dispersion is unstable). However, each mode also has one upper and one lower branch, which extend above and below its vertical location. In Fig. \ref{fig:dt0inst}, it can be seen that although each mode is stable ($\omega_i = 0$) by itself for small $\Delta x$, $\omega_i \neq 0$ for sufficiently large $\Delta x$. This is due to the overlapping of branches from adjacent modes as the vertical locations of the alias modes are lowered. Physically, such overlapping can result in unstable modes, for which $\omega_i > 0$ (or $\omega_i < 0$) corresponds to a growing (or damped) mode. However, one should distinguish the instability caused by overlapping of the branches of parallel aliases modes from that caused by the intersection of non-parallel modes, such as in the classic plasma two stream instability. 

Fig. \ref{fig:dt0inst} (b), (d), and (f) indicate that the instability first develops and its growth rate remains largest in magnitude near $k \Delta x = \pm \pi$ as $\Delta x$ increases. This can be understood from Fig.   \ref{fig:dt0inst} (a), (c), and (e), of $\omega_r$, which show that the branches of the $q=0$ mode, i.e., the finite grid size E.S.-like mode, $\omega_r = \omega_{ES}^\pm$, and the branches of $q=\mp 1$ aliaes  first intersect at $k \Delta x = \pm \pi$ for $\Delta x > \approx 0.5625$.  A second instability near $k \Delta x \approx 0$ is observed in Fig. \ref{fig:dt0inst} (f) when comparing Fig. \ref{fig:dt0inst} (d) and (f), although the associated instability is not as significant.



As discussed above, Fig.  \ref{fig:dt0inst} also indicates that there exists a threshold grid spacing, above which an unstable mode will be present. To obtain this threshold grid spacing, we can approximate $\omega_{ES}^\pm \approx \pm\text{sinc}^2\left(k \Delta x/2\right)/\gamma^{3/2}$ by keeping only the $q=0$ term in Eq. (\ref{eq:dt0}) and analytically solving for $\omega_t$. Since the solution $\omega_t$ is anti-symmetric with respect to $k \Delta x$, without loss of generality, we can consider the intersection of $\omega_{ES}^-$ with the $q = 1$ alias where the $q = 0, 1$ terms will dominate the sum in Eq. (\ref{eq:dt0}). Therefore, to a good approximation, $\omega_t$ satisfies, 

\be
1- \frac{16\Delta x^2}{\gamma^3}\sin^4{\frac{k\Delta x}{2}} \left[ \frac{1}{\omega_t^2 k^4 \Delta x^6} + \frac{1}{\left(2\pi-k\Delta x\right)^4\left(2\pi v_0+\omega_t \Delta x\right)^2}\right] \approx 0. \,\, (\omega_r<0)
\label{eq:terms2}
\ee

\begin{figure}[!htb]
  \centering
    \includegraphics[width=1\textwidth]{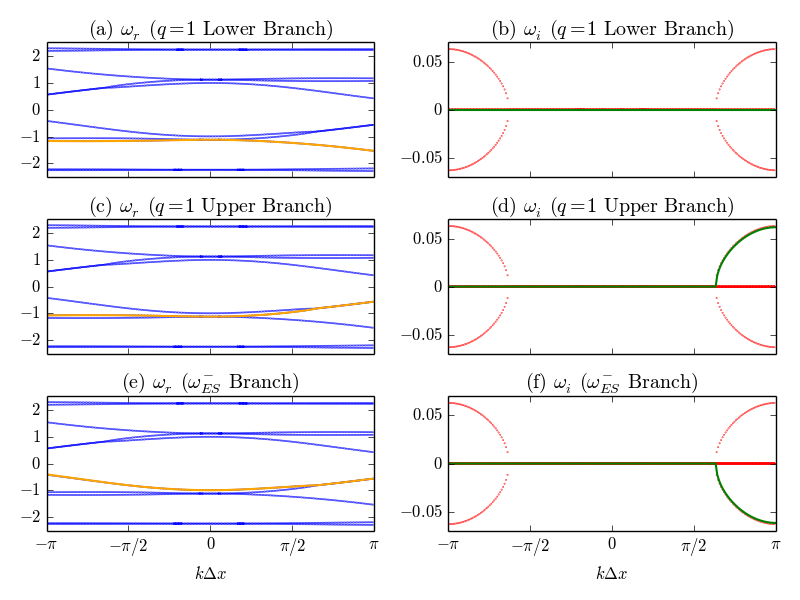}
    \caption{Analytic solutions to Eq. (\ref{eq:terms2}) (orange, green) and numerical solutions to Eq. (\ref{eq:dt0}) (blue, red) for $v_0=0.1$ and $\Delta x = 0.5625$. Plots in the left column show real parts of solutions, and plots in the right column show imaginary parts. All numerical roots are present in each set of plots, while each of the three analytic roots from Eq. (\ref{eq:terms2}) are shown in one row, respectively. The plots are labeled according to the analytic root shown. (a) and (b) Stable branch of the $q=1$ alias. (c) and (d) Unstable branch of the $q=1$ alias. (e) and (f) Lower branch of the finite difference E.S.-like mode. 
    }
    \label{fig:dt0approx}
\end{figure}

It is possible to solve Eq. (\ref{eq:terms2}) analytically for $\omega_t$. Unfortunately, the solutions to Eq. (\ref{eq:terms2}) are very complicated. They are of little practical use thus not given here. However,  a comparison of the analytic solutions of the approximated dispersion Eq. (\ref{eq:terms2}) against the numerical solutions of the exact dispersion Eq. (\ref{eq:dt0exanded}) presented in plots (c) and (d) of Fig. \ref{fig:dt0inst} is shown in Fig. \ref{fig:dt0approx}, for $v_0=0.1$ and $\Delta x = 0.5625$, to demonstrate the accuracy of the analytic solutions. There are three solutions to Eq. (\ref{eq:terms2}), i.e., one stable mode shown in Fig. \ref{fig:dt0approx} (a) and (b), and two unstable modes: one growing mode in Fig. \ref{fig:dt0approx} (c) and (d), and one damped mode in Fig. \ref{fig:dt0approx} (e) and (f).

Fig. \ref{fig:dt0approx} (d) and (f) show that the unstable modes are the upper branch of the $q = 1$ alias, and the $\omega_{ES}^-$ branch of the finite grid size plasma mode. The lower branch of the $q = 1$ alias in Fig. \ref{fig:dt0approx} (a) is predicted to be stable. Since both $\omega_r$ and $\omega_i$ plots are anti-symmetric with respect to $k\Delta x$ and the upper branch of the $q = 1$ alias in Fig. \ref{fig:dt0approx} (c) is associated with the upper right lobe in Fig. \ref{fig:dt0approx} (d), the lower branch of the $q=-1$ alias will then be associated with the lower left lobe of $\omega_i$ and is a damped mode. Similarly,  the imaginary part of the $\omega_{ES}^+$ branch corresponds to the upper left lobe in the $\omega_i$ plots in Fig. \ref{fig:dt0approx} and \emph{is} a growing mode, while the upper branch of the $q=-1$ alias is a stable mode. Solving the $q =0,-1$ analog of Eq. (\ref{eq:terms2}) indeed verifies these observations.


As we are most interested in when an instability occurs and its peak growth rate, solving Eq. (\ref{eq:terms2}) evaluated at $k \Delta x = \pm \pi$ gives,

\be
\omega_t\left(k \Delta x= \pm \pi\right) =
 \begin{cases}
-\frac{\pi  v_0}{\Delta x} \pm \sqrt{ 16 \pi^{-4} \gamma^{-3}  + (\frac{\pi  v_0}{\Delta x})^2 + 8  \pi^{-2} \gamma^{-3} \sqrt{4\pi^{-4} +(\frac{\pi  v_0}{\Delta x})^2  \gamma^3 }}\\
-\frac{\pi  v_0}{\Delta x} \pm \sqrt{ 16 \pi^{-4} \gamma^{-3}  + (\frac{\pi  v_0}{\Delta x})^2 - 8  \pi^{-2} \gamma^{-3} \sqrt{4\pi^{-4} +(\frac{\pi  v_0}{\Delta x})^2  \gamma^3 }}\\
 \end{cases}.
 \label{eq:thresholdsolution}
\ee

We therefore conclude that the maximum $\Delta x$ that will result in completely real solutions to Eq. (\ref{eq:dt0}) (i.e., for the term inside the square root of the second expression in Eq. (\ref{eq:thresholdsolution}) to be non-negative) is, 

\be
\Delta x_{\text{th}} = \frac{\pi^3\gamma^ {3/2} v_0}{4\sqrt{2}}.
\ee
For larger $\Delta x$, the most unstable mode has an imaginary part with magnitude 

\be
\left|\omega_i\left(k\Delta x =\pm\pi\right)\right|= \sqrt{\left|  16 \pi^{-4} \gamma^{-3}  + (\frac{\pi  v_0}{\Delta x})^2 - 8  \pi^{-2} \gamma^{-3} \sqrt{4\pi^{-4} +(\frac{\pi  v_0}{\Delta x})^2  \gamma^3 } \right|}.
\ee

The predicted threshold grid spacing compares well to the results obtained from numerically solving Eq. (\ref{eq:dt0}) using a "divide and conquer" method to determine the maximum stable grid spacing. This is illustrated in Table \ref{table:thresh}.

\begin{table}[!htbp]
\centering
\begin{tabular}{c c c}
\hline\hline
$v_0$ & $\Delta x_{\text{th}}$ & $\Delta x_-$ \\ [0.5ex] 
\hline
1/20 & 0.2746 & 0.2741 \\ 
1/10 & 0.5523 & 0.5517 \\
3/20 & 0.8363 & 0.833 \\
1/5 & 1.130 & 1.128 \\
1/4 & 1.438 & 1.435 \\
1/2 & 3.401 & 3.396 \\
$\sqrt{3}/2$ & 13.43 & 13.40 \\
9/10 & 17.14 & 17.12 \\
95/100 & 29.84 & 29.82 \\
975/1000 & 51.02 & 50.98 \\
999/1000 & 579.2 & ~578.8 \\ [1ex]
\hline
\end{tabular}
\caption{Comparison of the analytic and numerically determined threshold grid spacing for the 1D dispersion when $\Delta t \rightarrow 0$. The maximum grid spacing that does not result in an instability is predicted to be $\Delta x_\text{th}$. The largest grid spacing resulting in completely real solutions,  found from numerically solving Eq. (\ref{eq:dt0}), is given by $\Delta x_-$.}
\label{table:thresh}
\end{table}

\subsection{Instability from the full finite difference Electrostatic-like mode}

The relationship between the parameters $\Delta x$ and $\Delta t$ of Eq. (\ref{eq:esDisp}) and the numerical instability is investigated in this section for $\Delta t>0$. Because of the discretization in both space and time, all solutions of Eq. (\ref{eq:disp}) for $\omega$ in each spatial and temporal Brillouin zone should be identical. We again define $\omega_t=\omega-k v_0=\omega_r+i\omega_i$, and shall focus on the zeroth zone, i.e., the zone located in $\omega_r\Delta t\in\left[-\pi,\pi\right]$ (note that this is shifted from the zeroth $\omega \Delta t$ Brillouin zone, but does not affect our analysis since $\omega \Delta t$ Brillouin zones are periodic) and $k\Delta x\in \left[-\pi,\pi\right]$. Not surprisingly, the finite $\Delta t$ leads to significantly more complex numerical instability analysis, mainly due to the interaction of a $\omega=0$ mode and other modes in the system and the temporal Brillouin zones that change the location of the alias modes compared to the $dt = 0$ case.  Multiple instabilities can exist simultaneously for a particular set of parameters, for which we will focus on the \emph{leading} instability with the largest growth rate.

When examining the behavior of solutions in the zeroth zone, we shall further limit our discussion to those with $\omega_r>0$, since solutions with $\omega_r<0$ are just antisymmetric in $k\Delta x$. Note that solutions for $\omega_r\Delta t$ to Eq. (\ref{eq:disp}) in neighboring spatial Brillouin zones will be shifted by $\pm2\pi v_0\Delta t/\Delta x$, dependent on whether the Brillouin zone to the left ($+$) or right ($-$) is considered. This is again a consequence of searching for solutions to $\omega_t$, instead of $\omega$.

\subsubsection{Approximation of the dispersion by truncation}

Evaluating the infinite sum in Eq. (\ref{eq:esDisp}) exactly is difficult, 
so instead we multiply Eq. (\ref{eq:esDisp}) by $\gamma^3 \csc^4(k \Delta x/2)/(4 S \Delta x)$, truncate the infinite sum at the $N$th term and make the approximation,
\be
\epsilon_a^{N} \approx 0
\label{eq:dxdtApprox}
\ee
with
\begin{equation}
\begin{split}
\epsilon_a^{N} &= \frac{\gamma^3\text{csc}^4\left(k\Delta x/2\right)}{2 S^2\Delta x^2} \sin{\left [ \left(S v_0 k\Delta x+\omega_t \Delta t\right)/2\right ] } \\
&- \sum_{q=-N}^{N}  \frac{2}{\left(k\Delta x-2 \pi q\right)^4}\text{csc}\left(\pi q S v_0 + \omega_t\Delta t/2\right)\\
 &- \sum_{q=-N}^{N}  \frac{S v_0}{\left(k\Delta x-2 \pi q\right)^3}\text{csc}^2\left(\pi q S v_0 + \omega_t\Delta t/2\right)\cos\left(\pi q S v_0 + \omega_t\Delta t/2\right),
\end{split}
\nonumber
\end{equation}
where $S=\Delta t /\Delta x$ is the Courant number. For the FDTD Yee solver to be stable, $S < 1$.


There are, however, unavoidable effects of approximating the sum in Eq. (\ref{eq:esDisp}) with any finite sum. One effect is that a solution that is from a truncated term and responsible for an unstable mode will not be present over some parameter ranges. This effect will be most significant when the alias associated with such term are responsible for the leading, or near leading, instability. However, the leading instability is usually due to the first few terms in the sum, as further terms decrease at least as fast as $1/(k\Delta x - 2 \pi q)^3$. This consideration justifies a truncation of the infinite sum at a relatively small $N$. Another unavoidable effect is that solutions across all Brillouin zones will not be symmetric. These effects are artifacts of the approximation and not actually properties of Eq. (\ref{eq:esDisp}). 

\begin{figure}[ht]
  \centering
    \includegraphics[width=1\textwidth]{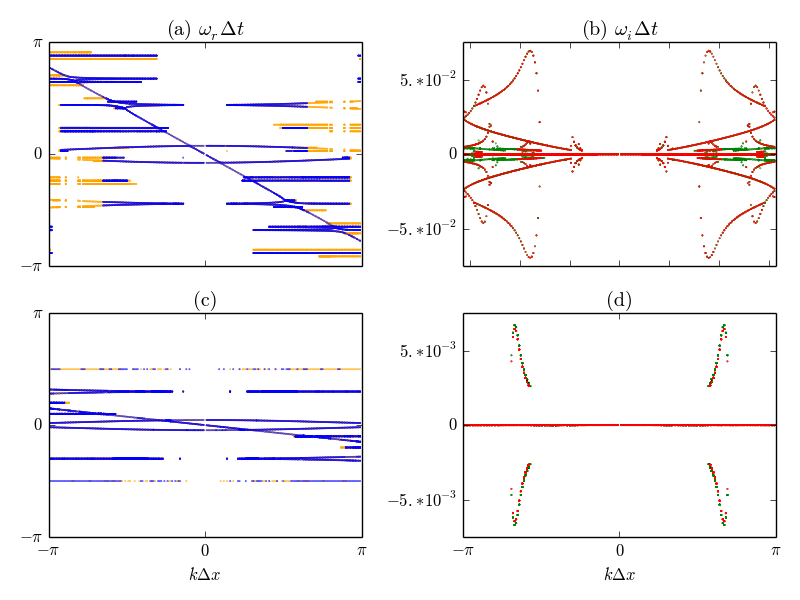}
    \caption{Top: Solutions to Eq. (\ref{eq:dxdtApprox}) with $N=10$ (Blue, Red) and $N = 20$ (Orange, Green). Top: $S = 9/10$, $v_0=\sqrt{3}/2$, $\Delta x = 3/4$. Bottom: $S = 3/5$, $v_0=1/4$, $\Delta x = 1/4$. Plots (a), (c) show solutions of $\omega_r\Delta t$, and plots (b), (d) show solutions of $\omega_i\Delta t$. Note that including more terms to the sum in Eq. (\ref{eq:dxdtApprox}) necessarily introduces solutions at more points (i.e. the orange and green curves are present "underneath" the blue and red curves.)}
    \label{fig:compPlts}
\end{figure}

For the remainder of this section, we will adopt Eq. (\ref{eq:dxdtApprox}) with $N\rightarrow 10$ as our standard. We will see that this provides the dominant behavior of Eq. (\ref{eq:esDisp}) for $1/10\lessapprox \sS $, where we have introduced the scaled Courant number $\sS = S v_0$.
Fig. \ref{fig:compPlts} shows the approximate convergence of solutions for $\omega_t\Delta t$ with some specific parameter sets.

Fig. \ref{fig:compPlts} indicates that, in addition to the $q=0$ finite grid size E.S.-like mode and $|q|>0$ alias modes similar to the $\Delta t \rightarrow 0$ case, there is a negative-slope linear mode that passes through the origin. This mode corresponds to the second $\omega = 0$ mode that could not be factored from $\epsilon_a$, as was possible in Eq. (\ref{eq:dt0}) for the $\Delta t \rightarrow 0$ case. This mode is given approximately by $\omega_{t0}\Delta t \approx - k\Delta x \sS$. Furthermore, the vertical locations of the $|q|>0$ alias modes are no longer necessarily proportional to $q$ as these modes need to be folded into the zeroth Brillouin zone. With these differences, one can expect instability to occur in the following two scenarios,

\begin{enumerate}
\item The finite grid size E.S.-like mode and aliased modes intersect each other. Because we are interested only in the leading instability, we will analyze the intersection of the finite grid size E.S.-like mode and $q=\pm1$ modes in detail and note that this is of concern only over limited paramater ranges.
\item Alias modes intersect the $\omega_{t0}$ mode (see Fig. \ref{fig:compPlts}). This causes instability across a significant portion of the parameter ranges and is often responsible for the leading instability. 
\end{enumerate}

To investigate these two types of instabilities, we will first determine the location of the $|q|>0$ alias modes. Similar to the $\Delta t \rightarrow 0$ case, in the non-relativistic limit (i.e. $v_0 \ll 1$) and for sufficiently small $q$, the location of the $q^{th}$ ($q\neq 0$) alias is determined by  $\omega_t \rvert^q= - 2\pi q v_0/\Delta x$ or $\omega_t\Delta t\rvert^q = -2\pi q \sS$. The relativistic generalization can be found from the pole, $\sin \left(\pi q \sS + \omega_t\Delta t/2\right) = 0$, in the $q^{th}$ term in Eq. (\ref{eq:dxdtApprox}). The solution is $ \omega_t\Delta t \rvert^q = 2\pi (n- q \sS) $, where $n $  is an integer and $ n \in [q\sS -1/2, q\sS +1/2) $ for $q>0$, $ n \in (q\sS -1/2, q\sS +1/2] $ for $q<0$, as $\omega_t\Delta t \rvert^q \in [-\pi, \pi]$.


The location of the alias mode deserves some further discussion. When $2\sS = l/m$, where $l, m$ are coprime integers, it can be shown that there are only $2m+1$ possible locations equally spaced between $\omega_t \Delta t = -\pi$ and $\omega_t \Delta t = \pi$ (including $-\pi$ and $\pi$) for the alias modes. For example, when $\sS = 1/2$, all $q>0$ aliases will be located at $\omega_t \Delta t \rvert^{q}= \left\{ -\pi, 0 \right\} $ (although the exact location of the alias branches may be somewhat different than these two values). The $q<0$ modes will be at symmetric locations between $0$ and $\pi$. Similarly, for $\sS= \left\{ 1/4,3/4 \right \}$, we find $\omega_t \Delta t\rvert^{q} = \left \{ 0, \pm\pi/2,\pm\pi \right \}$. It has been observed that for these exact $\sS$ values no instability is present. For other $\sS$ values, although many aliases may overlap at the same locations, this type of intersection may not be responsible for the leading instability. This is because the index $q_1$ and $q_2$ of two different alias modes having the same location satisfy $|q_1 - q_2| = |(n_1 - n_2)/\sS| \ge \sS^{-1}$. For $\sS \ll 1$, $|q_1 - q_2| \gg 1$, therefore the contribution of the high $q$ alias mode in the infinite sum is always highly suppressed. 


\subsubsection{Instability from $q=0,1$ modes}

In the case that $\sS$ is sufficiently small, the $q=1$ alias mode is near the finite space-time E.S.-like ($q=0$) mode. It also has branches going above and below $\omega_t\Delta t\rvert^{1+}$. These two branches can be understood by recalling that solutions to Eq. (\ref{eq:disp}) in each spatial and temporal Brillouin zone should be identical (up to a phase shift $2\pi v_0/\Delta x$ in $\omega$, i.e., $2\pi \sS $ in $\omega_t \Delta t$). In the $1^{st}$ spatial Brillouin zone, the $q = 1$ alias has identical shape to the $q = 0$ mode of the zeroth zone. Therefore, one can calculate the branch separation of the $q = 1$ alias at $k \Delta x = \pi$ by the branch separation of the $q=0$ mode at $k \Delta x = \pi$.

The branch separation of the $q=0$ mode at $k\Delta x = \left\{ 0, \pm \pi \right \}$ can be found by only including the $0^{th}$ term from the sum in Eq. (\ref{eq:dxdtApprox}). At $k\Delta x = 0$, we obtain
\be
 2\gamma^3 \left[ \text{cos}\left(\omega_t\Delta t\right)-1 \right] +S^2\Delta x^2  =0,
\ee
which gives where the finite space-time E.S.-like solutions crosses the $\omega_t\Delta t$ axis: $\omega_t\Delta t\rvert^{0\pm} \allowbreak (k\Delta x=0)= \pm\text{arccos}\left(1-S^2\Delta x^2/2\gamma^3\right)$. Therefore, the branch separation of this mode at $k\Delta x=0$ is $2\, \text{arccos}\left(1-S^2\Delta x^2/2\gamma^3\right)$. When $S \Delta x /\gamma^{3/2} \ll 1$, $\omega_t \Delta t\rvert^{0\pm} (k \Delta x =0) \approx \pm S \Delta x /\gamma^{3/2} = \pm \Delta t / \gamma^{3/2}$ and the branch separation is about $2S \Delta x /\gamma^{3/2} = 2\Delta t / \gamma^{3/2}$.

At $k\Delta x = \pm \pi$, we obtain, 
\be
-\frac{8 S \Delta x  \csc (\omega \Delta t /2) }{\pi ^4 \gamma^3}-\frac{2 S^2 \Delta x  v_0 \csc^3 (\omega \Delta t) \sin(\omega \Delta t)}{\pi ^3 \gamma^3}+\frac{2 \sin [(\pi  S \text{v0}+ \omega \Delta t)/2 ] }{S \Delta x}  =0,
\label{eq:mode0pi}
\ee
which can be solved analytically to give $\omega_t\Delta t\rvert^{0\pm} (k\Delta x= \pm \pi)$ and the branch separation there. But the result is again very complicated and not given here. Instead we present an approximate yet simple prediction of the instability from $q=0, \pm1$ modes below.  

\begin{figure}[h!]
  \centering
   \includegraphics[width=1\textwidth]{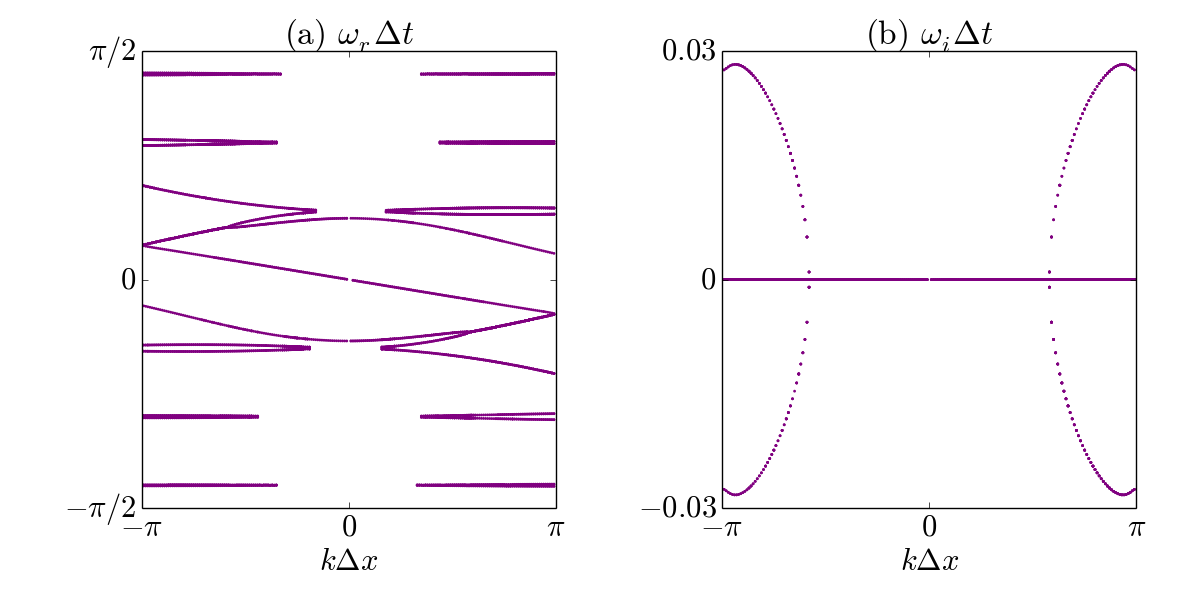}
    \caption{The $\delta$ instability for $\sS = 0.075$, $\Delta x = 0.5625$, $v_0 = 1/10$. (a) $\omega_r\Delta t$ , (b) $\omega_i\Delta t$. For these parameter values $\Delta x^\delta_{th}=0.6331$, and solutions to Eq. (\ref{eq:dxdtApprox}) in the zeroth zone are contained in the range of (a).}
    \label{fig:delta}
\end{figure}

Similar to the $\Delta t \rightarrow 0$ case, when the lower branch of the $q=1$ (or $q=-1$) alias mode and the $q=0$ mode intersect at $k\Delta x =  \pi$  (or $- \pi $), it causes an instability, as shown in Fig. \ref{fig:delta}. In Fig.  \ref{fig:delta} (b), the upper lobe near $k \Delta x = \pi$ is associated with the upper branch of the $q=1$ mode, and the lower lobe is from the lower branch of the $q=0$ alias mode.


The condition for these two modes to touch at $k \Delta x =  \pi$ can be approximately found by letting $|\omega_t\Delta t\rvert^{0} (k\Delta x= 0)| = |\omega_t\Delta t\rvert^{1\pm} (k\Delta x= 0)|$, i.e. $\Delta t / \gamma^{3/2} = 2\pi \sS$, which gives $\Delta x_{th}^{\delta} = 2 \pi v_0 \gamma^{3/2}$, where the superscript $\delta$ is used to denote the threshold for this type of mode intersection. When $\Delta x$ is close to or larger than this value, the intersection of the $q=0$ and $q=\pm1$ modes give rise to the instability shown in Fig.  \ref{fig:delta}, which we name  the $\delta$-instability. For $\Delta x < \approx \Delta x_{th}^{\delta}$, these two modes detach and are stable. 

However, due to the folding of high $q$ alias modes into a lower location than the $q=1$ mode, it is possible that such mode can interact with the $q=0$ mode causing instability. The growth rate for the instability in this case usually does not lead as it involves a high $q$ mode. The threshold of the $\delta$-instability is not very restrictive as typical simulation grid size satisfies $\Delta x < \Delta x_{th}^{\delta}$ for sufficiently large $v_0$. However, as $\Delta x$ decreases, the $q=\pm 1$ mode will be folded into the opposite half of the zeroth Brillouin zone and intersect with $\omega = 0$ mode which is discussed in the next section. 

\subsubsection{Instability from the interaction between $\omega=0$ mode and alias modes}

It can be shown that $q=0, 1$ modes first intersect at $k \Delta x = \pi$ and $\omega_t \Delta t = -\pi \sS$ as the separations of the branches of each mode are equal at $k \Delta x = \pi$. We note that $k \Delta x = \pi$ and $\omega_t \Delta t = -\pi \sS$ is also on the $\omega_{t_0}$ mode, i.e., the $\omega_{t_0}$ mode also intersects the $q=0, 1$ modes at the same location. It has been observed that the intersection of the $\omega_{t_0}$ mode with the $q=0$ mode is always stable, while the intersection with $|q|>0$ alias modes is always unstable.    


We devote the remainder of this section to analyzing the instability resulting from $\omega_{t0}$-alias intersections, since this type of instability is by far the most prevalent in the parameter range $v_0 \lesssim 1$ of our interest.

To obtain a simple expression for the peak growth rate that results from an intersection of the $q^{th}$ alias with $\omega_{t0}$, we will approximate the solution to Eq. (\ref{eq:disp}) in the vicinity of $\omega_t\Delta t\rvert^{q\pm}$ by including only the $q^{th}$ term in the sum. Therefore, we can solve
\be
\begin{split}
\frac{\gamma^3\text{csc}^4\left(k\Delta x/2\right)}{2 S^2\Delta x^2}&\sin{\left\{\left(S v_0 k\Delta x+\omega_t \Delta t\right)/2\right\}}
-\frac{2}{\left(k\Delta x-2 \pi q\right)^4}\text{csc}\left(\pi q S v_0 + \omega_t\Delta t/2\right)\\
 -&\frac{S v_0}{\left(k\Delta x-2 \pi q\right)^3}\text{csc}^2\left(\pi q S v_0 + \omega_t\Delta t/2\right)\cos\left(\pi q S v_0 + \omega_t\Delta t/2\right)=0
\end{split}
\label{eq:oneTerm}
\ee
exactly for $\omega_t\Delta t$.

\begin{figure}[h!]
  \centering
    \includegraphics[width=1\textwidth]{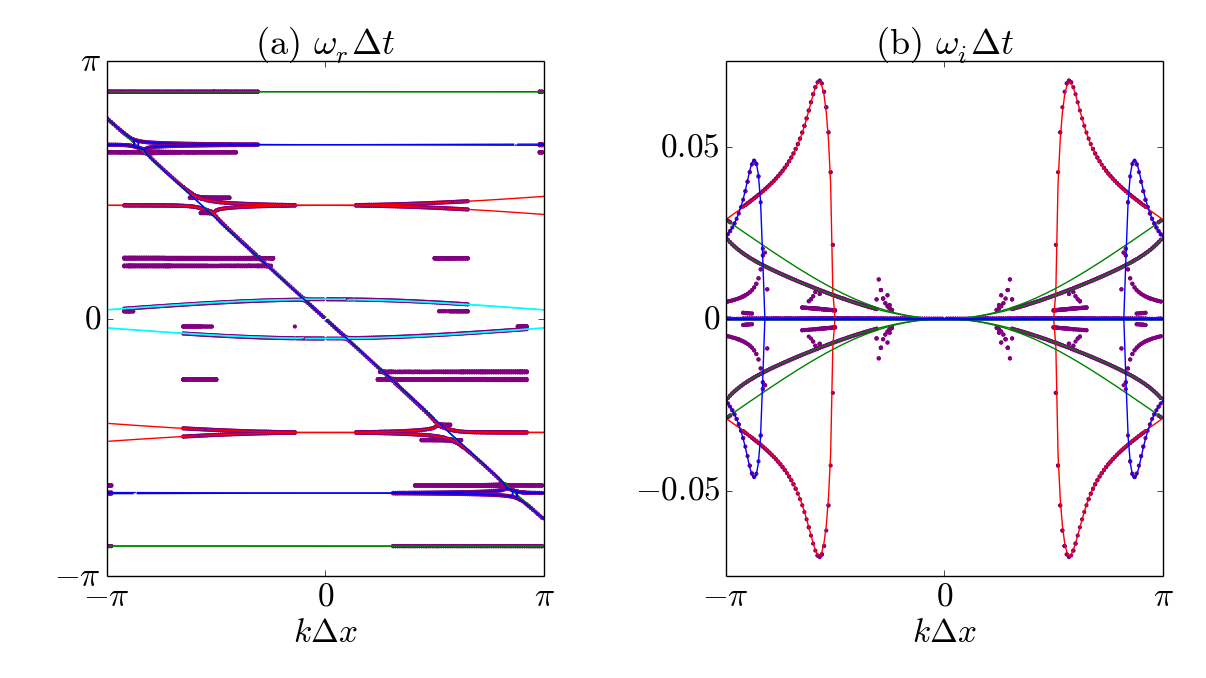}
    \caption{Approximated analytic solutions to Eq. (\ref{eq:oneTerm}) for the $q= 0$ (cyan), $\pm1$ (red), $\pm2$ (green), $\pm3$ (blue) aliases have been overlaid on the numerical solutions (dots) to Eq. (\ref{eq:dxdtApprox}) for $S = 0.9$, $v0=\sqrt{3}/2$, $\Delta x = 0.75$. (a) Solutions to $\omega_r\Delta t$. (b) Solutions to $\omega_i\Delta t$. }
    \label{fig:aliases}
\end{figure}

The analytic solutions to Eq. (\ref{eq:oneTerm}) shown in Fig. \ref{fig:aliases} are accurate for all $k\Delta x$, as long as there is no overlapping alias with smaller $|q|$ index. These solutions are also complicated. We focus on the $q>0$ alias and notice that the peak growth rate of this alias occurs near where it intersects the $\omega_{t0}$ mode. We then solve Eq. (\ref{eq:oneTerm}) with $k\Delta x= 2\pi\left(q\sS -n \right) / \sS$, where $n$ is an integer and $q\sS-1/2 \le n < q\sS + 1/2$. When $n=0$, the $q^{th}$ alias is at $\omega \Delta t = -2\pi q \sS$ and not folded into the upper Brillouin zone, while the $\omega_{t0} \in [-\pi \sS,  \pi \sS]$. Therefore, there is no interaction or instability between the $\omega_{t0}$ mode and alias modes for $n=0$ (i.e., when this mode is not folded). For $n \ne 0$, one can rewrite Eq. (\ref{eq:oneTerm}) as,

\be
1+c_1 \text{ cot}\left( \alpha \right)+c_2/\left[1+\text{ cot}^2 \left( \alpha \right) \right] =0
\label{eq:form}
\ee
where $c_1 = (k\Delta x/2 - \pi q) \sS =  -n\pi $, $c_2 = 4\left(-1\right)^{n+1}\gamma^3  \left[ S\Delta x\text{ sinc}^2 \left( n\pi/\sS \right) \right ]^{-2}$, and $\alpha = \pi q\sS +\omega_t\Delta t /2$. 


Since $n \ne 0$, the $q^{th}$ alias is folded and there are three solutions to Eq. (\ref{eq:form}) in the zeroth Brillouin zone. One solution for $\omega_t\Delta t$ is completely real, and two solutions have a nonzero imaginary part. The latter are of interest and are complex conjugates.   
%
%
For $|\omega_i \Delta t| \ll 1$, we find that
\be
\omega_i\Delta t \approx \frac{\left(-1\right)^{n+1}\sqrt{3}\left(n\pi\right)^{1/3}}{\gamma}\left[\text{sinc}^2\left( n \pi/\sS \right) S\Delta x/2\right]^{2/3}
\label{eq:growth}
\ee
Eq. (\ref{eq:growth}) allows us to predict the growth rate at the intersection of the $q^{th}$ alias with $\omega_{t0}$, when $q\sS \in\left(n-1/2,n+1/2\right]$.

For the $q<0$ alias modes, we solve an equation that is exactly the same as Eq. (\ref{eq:form}) to get the same peak growth rate from their intersections with the $\omega_{t0}$ mode.  This is due to the symmetry of $\pm q$ aliased solutions. We thus conclude that Eq. (\ref{eq:growth}) gives the peak growth rate for the intersection of the $q^{th}$ alias with $\omega_{t0}$ for $|q|>0$. Table \ref{table:inst} compares the instability predicted by Eq. (\ref{eq:growth}) to that obtained by solving Eq. (\ref{eq:dxdtApprox}) at the intersection point.

\begin{table}[ht]
\centering
\begin{tabular}{c c c c c}
 && &      & \\
\hline\hline
$q$ & $\omega_i\Delta t_{pred}$ & $\omega_i\Delta t_{solved}$ & $\omega_i\Delta t_{pred}$ & $\omega_i\Delta t_{solved}$ \\ [0.5ex] 
\hline
1&0.0684 & .0692 & -- & -- \\ 
2&0.0466 & 0.0460 & 0.00465& 0.00490  \\
3&0.0466 & 0.0465 & 0.00465&0.00490 \\
4&0.0112 & 0.0115 & 0.00465&0.00490 \\
5&0.00713 & 0.00732 & 0.00465&0.00490 \\
6&0.0183 & 0.0185 & 0.00498 &0.00502\\
7&0.0183 & 0.0185 & 0.00498& 0.00502 \\
8&0.0121 & 0.0062 & 0.00498 &0.00502\\
9&0.00032 & .00018 & 0.00498 &0.00502\\
10&0.0080 & 0.000 & 0.00427 &0.00427\\[1ex]
\hline
\end{tabular}
\caption{The growth rates predicted by Eq. (\ref{eq:growth}), $\omega_i\Delta t_\text{pred}$, compared to the growth rates found by solving Eq. (\ref{eq:dxdtApprox}) at the alias-$\omega_{t0}$ mode intersection point, $\omega_i\Delta t_\text{solved}$. The left two columns are for $S=0.9$, $\Delta x = 0.75$, and the right two columns are for $S=0.300$, $\Delta x = 1$. Both cases are for $v_0 = \sqrt{3}/2$.}
\label{table:inst}
\end{table}

\section{Summary}\label{sec:summary}

The PIC method is particularly useful for the study of the kinetic behavior of the plasma compared to other numerical plasma models, due to its first principles treatment of the particle dynamics and the self-consistency of the solution. PIC models, especially E.M. PIC models, are increasingly being applied to relativistic particle beam and plasma modeling. However, a fundamental incompatibility between the continuous particle model and the discrete field representation in PIC models causes the aliasing effect. 
Under certain numerical conditions, the alias modes can interact with the other modes admitted in the E.M. PIC model, causing numerical instabilities that can render unphysical simulation results or even destroy the simulation. 

In this paper, we review the numerical dispersion relation of the standard E.M. PIC algorithm based on the Yee FDTD solver for the analysis of the origin of the instabilities. We abide by the following two guidelines that have been generally overlooked in the past: (1) the root cause of aliasing is the sampling of a continuous variable onto a discrete variable, and the perturbation in a linear stability analysis should be applied to one or more internal variable of the system under investigation, so as to avoid artificial inclusion of the aliasing effect. (2) One should avoid making unnecessary algebraic  approximations that may result in an approximation to the original model. Based on these guidelines, we rigorously derive the faithful numerical dispersion of the standard E.M. PIC algorithm with a simple current deposition scheme. 

We analyzed the numerical dispersion of the E.S.-likes modes in a 1D E.M. PIC simulation with a drifting cold plasma. Accurate numerical solutions and corresponding approximated growth rates are obtained for the finite grid instability. We have shown that in the $\Delta t \rightarrow 0$ limit and for the parameter regime corresponding to $\Delta t >0, \Delta x > \Delta x_{th}^{\delta}$, the dominant finite grid instability is caused by the intersection of the $q=0$ mode and $q=\pm 1$ alias modes (Fig.  \ref{fig:dt0approx} and \ref{fig:delta}). On the other hand, for relativistic cold plasma flow under most relevant simulation parameters, an intersection of the $\omega = 0$ mode and alias modes can lead to the most dominant instability (Fig. \ref{fig:aliases}). Following the guidelines mentioned before allows us to predict the locations of the alias modes in the system and the conditions for numerical instabilities to occur. In a future work, we will extend this analysis to the E.M.-like modes and to multiple dimensions for better understanding of the PIC modeling of relativistic beam and plasma systems.

\bibliography{writeupV8}{}

\begin{thebibliography}{10}

\bibitem{Birdsall-Langdon}
{\sc C.~K. Birdsall} and {\sc A.~B. Langdon},
\newblock {\em {Plasma Physics via Computer Simulation}},
\newblock IOP Publishing Ltd., Bristol, England, 1991.

\bibitem{Hockney-Eastwood}
{\sc R.~W. Hockney} and {\sc J.~W. Eastwood},
\newblock {\em {Computer Simulation Using Particles}},
\newblock CRC Press, 1988.

\bibitem{Dawson1983}
{\sc J.~Dawson},
\newblock {\em Reviews of Modern Physics} {\bf 55}, 403 (1983).

\bibitem{Yee1966}
{\sc K.~S. Yee},
\newblock {\em IEEE Transactions on Antennas and Propagation} {\bf 14}, 302
  (1966).

\bibitem{Langdon1970}
{\sc A.~Langdon},
\newblock {\em Journal of Computational Physics} {\bf 6}, 247 (1970).

\bibitem{Vay2011}
{\sc J.-L. Vay}, {\sc C.~Geddes}, {\sc E.~Cormier-Michel}, and {\sc D.~Grote},
\newblock {\em Journal of Computational Physics} {\bf 230}, 5908 (2011).

\bibitem{Spitkovsky2008}
{\sc A.~Spitkovsky},
\newblock {\em The Astrophysical Journal} {\bf 673}, L39 (2008).

\bibitem{SamueldeFreitasMartins}
{\sc {S. F. Martins}},
\newblock {\em {Pushing the limits of computer simulations for
  ultra-relativistic scenarios}},
\newblock PhD thesis, INSTITUTO SUPERIOR TECNICO, Portugal, 2011.

\bibitem{Vay2007}
{\sc J.-L. Vay},
\newblock {\em Physical Review Letters} {\bf 98}, 130405 (2007).

\bibitem{Martins2010}
{\sc S.~F. Martins}, {\sc R.~a. Fonseca}, {\sc W.~Lu}, {\sc W.~B. Mori}, and
  {\sc L.~O. Silva},
\newblock {\em Nature Physics} {\bf 6}, 311 (2010).

\bibitem{GODFREY1974}
{\sc B.~B. GODFREY},
\newblock {\em Journal of Computational Physics} {\bf 15}, 504 (1974).

\bibitem{Godfrey1975}
{\sc B.~B. Godfrey},
\newblock {\em Journal of Computational Physics} {\bf 19}, 58 (1975).

\bibitem{Godfrey2013}
{\sc B.~B. Godfrey} and {\sc J.-L. Vay},
\newblock {\em Journal of Computational Physics} {\bf 248}, 33 (2013).

\bibitem{Xu20132503}
{\sc X.~Xu}, {\sc P.~Yu}, {\sc S.~F. Martins}, {\sc F.~S. Tsung}, {\sc V.~K.
  Decyk}, {\sc J.~Vieira}, {\sc R.~A. Fonseca}, {\sc W.~Lu}, {\sc L.~O. Silva},
  and {\sc W.~B. Mori},
\newblock {\em Computer Physics Communications} {\bf 184}, 2503  (2013).

\bibitem{Villasenor1992}
{\sc J.~Villasenor} and {\sc O.~Buneman},
\newblock {\em Computer Physics Communications} {\bf 69}, 306 (1992).

\bibitem{Esirkepov2001}
{\sc T.~Esirkepov},
\newblock {\em Computer Physics Communications} {\bf 135}, 144 (2001).

\bibitem{Greenwood2004}
{\sc A.~D. Greenwood}, {\sc K.~L. Cartwright}, {\sc J.~W. Luginsland}, and {\sc
  E.~A. Baca},
\newblock {\em Journal of Computational Physics} {\bf 201}, 665 (2004).

\bibitem{McKinstrie1995}
{\sc C.~J. McKinstrie} and {\sc E.~A. Startsev},
\newblock {\em Physics of Plasmas} {\bf 2}, 3234 (1995).

\end{thebibliography}
\bibliographystyle{phjcp}



\end{document}